\documentclass[%
 reprint,
groupedaddress,
nofootinbib,
 amsmath,amssymb,
 aps,
prb,
]{revtex4-2}

\usepackage{graphicx}
\usepackage{dcolumn}
\usepackage{bm}
\usepackage{dsfont}
\usepackage{amsmath}
\usepackage{braket}
\usepackage{courier}
\usepackage{appendix}
\usepackage[normalem]{ulem}
\usepackage{graphicx}
\usepackage{cancel}
\usepackage{printlen}
\usepackage[caption=false]{subfig}
\usepackage{xcolor}
\usepackage[normalem]{ulem}
\usepackage[colorlinks=true,
    linkcolor=blue,
    citecolor=red,
    filecolor=magenta,      
    urlcolor=cyan,
    pdftitle={Honeycomb},]{hyperref}
\usepackage{pgfplots}
\usepackage{tikz}
\usepackage{float}
\usepackage{comment}

\newcommand{\matteo}[1]{\textsf{\color{green!70!black}(MI: #1)}}

\begin{document}

\title{Topology, criticality, and dynamically generated qubits in a stochastic measurement-only Kitaev model} 

\author{Adithya Sriram}
\author{Tibor Rakovszky}
\author{Vedika Khemani}
\author{Matteo Ippoliti}
\affiliation{%
 Department of Physics, Stanford University, Stanford, California 94305, USA
}%

\date{\today}

\begin{abstract}
We consider a paradigmatic solvable model of topological order in two dimensions, Kitaev's honeycomb Hamiltonian, and turn it into a measurement-only dynamics consisting of stochastic measurements of two-qubit bond operators. 
We find an entanglement phase diagram that resembles that of the Hamiltonian problem in some ways, while being qualitatively different in others. 
When one type of bond is dominantly measured, we find area-law entangled phases that protect two topological qubits (on a torus) for a time exponential in system size. This generalizes the recently-proposed idea of Floquet codes, where logical qubits are dynamically generated by a time-periodic measurement schedule, to a stochastic setting.
When all types of bonds are measured with comparable frequency, we find a critical phase with a logarithmic violation of the area-law, which sharply distinguishes it from its Hamiltonian counterpart. The critical phase has the same set of topological qubits, as diagnosed by the tripartite mutual information, but protects them only for a time polynomial in system size. 
Furthermore, we observe an unusual behavior for the dynamical purification of mixed states, characterized at late times by the dynamical exponent $z = 1/2$---a super-ballistic dynamics made possible by measurements.
\end{abstract}

\maketitle


\section{Introduction}

Understanding how entanglement is generated and stabilized in many-body systems, and particularly {\it open} systems, is an important direction for both quantum information science and physics~\cite{amico_entanglement_2008, horodecki_quantum_2009, abanin_colloquium_2019}.
Recently it was realized that, for a specific kind of open quantum system (one whose interaction with the environment is represented by {\it measurements} with known outcomes), the dynamics may spontaneously generate and preserve large amounts of entanglement. Even more strikingly, this phenomenon defines a new kind of phase structure far from equilibrium, with phases identified by the structure of quantum entanglement in late-time states~\cite{skinner_measurement-induced_2019, li_quantum_2018, li_measurement-driven_2019, chan_unitary-projective_2019, gullans_dynamical_2020, choi_quantum_2020, potter_entanglement_2021, szyniszewski_universality_2020}:
upon tuning the rate or strenght of measurements, the system transitions from a phase with limited entanglement (an {\it area-law} phase~\cite{eisert_colloquium_2010}) to one with extensive entanglement (a {\it volume-law} phase), separated by a sharp phase transition~\cite{zabalo_critical_2020, li_conformal_2021}.
These entanglement phases can be enriched with additional structure, such as symmetry~\cite{ippoliti_entanglement_2021, lavasani_measurement-induced_2021, sang_measurement_2020, bao_symmetry_2021, li_robust_2021, agrawal_entanglement_2021, barratt_field_2021} and { topology}~\cite{lavasani_topological_2021, ippoliti_entanglement_2021}.

A heuristic understanding of the entanglement phase transition is based on a competition between ``scrambling'' due to unitary dynamics and the disentangling effect of projective measurements. 
However, this picture is incomplete. In fact, measurements alone can stabilize an entangling phase~\cite{ippoliti_entanglement_2021}. 
In such {\it measurement-only} dynamics, the principle governing the entanglement phases and transitions is not a rate (or strength) of measurements, but rather the structure of the operators being measured.

Topology can naturally be introduced in this setting by considering {\it commuting projector} models of topological order, such as the toric code Hamiltonian~\cite{kitaev_fault-tolerant_2003}. 
Frequent measurements of the (commuting) terms in the topological Hamiltonian may win over sufficiently infrequent measurements of ``trivial'' terms, and stabilize an area-law entangled phase with topological order~\cite{lavasani_topological_2021, ippoliti_entanglement_2021}.
This setting is closely related to {\it stabilizer quantum error correction}~\cite{shor_scheme_1995, gottesman_class_1996}, wherein a set of commuting operators (the stabilizers) are repeatedly measured to glean information about errors (e.g., unwanted interactions with the environment) that may have taken place in the system.
The stabilizers are identified with the commuting Hamiltonian terms, and the topological phase is thus identified with an error-correcting phase, where ``errors'' (measurements of trivial operators) fail to propagate and damage the information encoded in the topological ground space.

It is interesting to ask whether the possibility of topological order in monitored dynamics survives in more general models, away from an explicit commuting-projectors limit.
To this end, it is useful to consider another paradigm for quantum error correction, given by {\it subsystem codes}~\cite{poulin_stabilizer_2005, bacon_operator_2006, aliferis_subsystem_2007}. 
Unlike stabilizer codes, these consist of measurements (``checks'') that need not commute, and thus cannot be measured simultaneously. However, upon measuring the checks sequentially, one can still learn about errors on the system and potentially correct them. (In practice, a reason to consider such codes is that the operators to be measured are often smaller.) In a measurement-only dynamics based on subsystem codes, the non-commutativity of checks may give rise to interesting dynamics even in the absence of ``errors'', i.e. additional measurements of trivial operators or unitary operations, and potentially stabilize topological order (in the form of encoded logical qubits) even in a stochastic setting.

A closely related recent development is the discovery of {\it Floquet codes}~\cite{hastings_dynamically_2021, gidney_fault-tolerant_2021, haah_boundaries_2022, aasen_adiabatic_2022, paetznick_performance_2022, vuillot_planar_2021}. These are, in essence, measurement-only dynamics that follow a regular, time-periodic schedule, and in doing so stabilize one or more logical qubits. Crucially, however, {\it no} logical qubit would exist if the specific time-dependence were not imposed, i.e., if all the allowed measurements were viewed as checks for a subsystem code. In this sense, the logical qubits are ``dynamically generated''. 
The original Floquet code was defined on a honeycomb lattice~\cite{hastings_dynamically_2021}, based on the celebrated Kitaev Hamiltonian model of an exactly-solvable spin liquid~\cite{kitaev_anyons_2006}. 
While the specific time-periodic schedule is of practical interest (as it allows a fault-tolerant decoding protocol), the dynamical generation of topological qubits is a remarkable phenomenon unto itself, and is interesting to study more generally, including in the framework of stochastic measurement-only dynamics.

\begin{figure*}
    \centering
    \includegraphics[width=\textwidth]{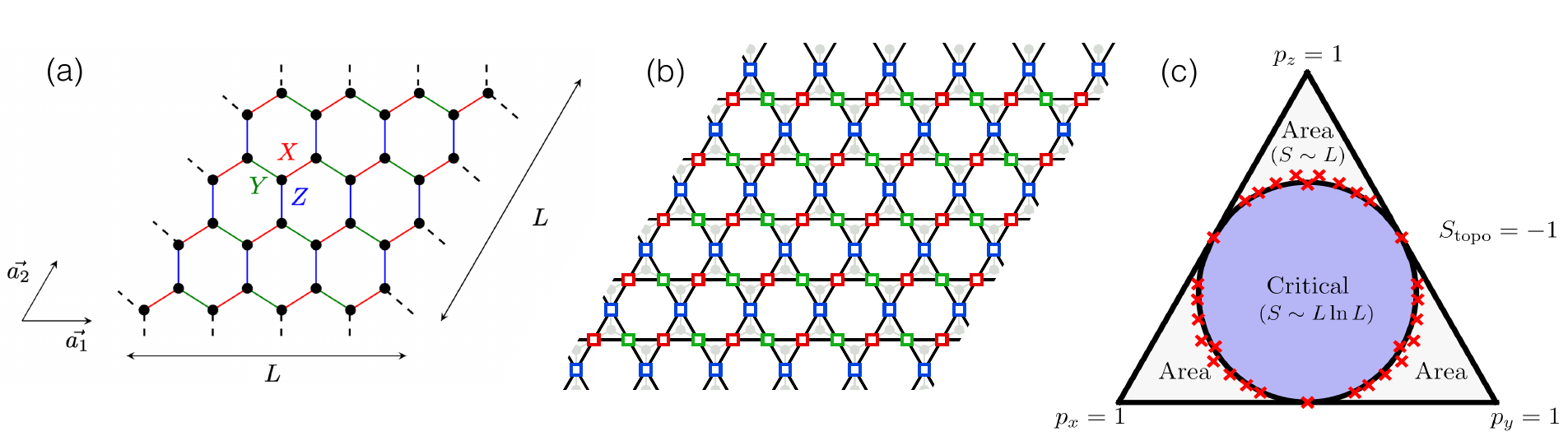}
    \caption{(a) Honeycomb layout with $L = 4$ two qubit basis cells in each direction. There are $2L$ qubits in each direction. (b) Frustration graph of the honeycomb model. Vertices (squares) represent two-qubit operators (color-coded like the bonds in (a)), edges represent anticommutation. The honeycomb lattice of physical qubits (light gray) is also shown as a guide to the eye. (c) Phase diagram of measurement-only circuit model. The red $x$'s show approximate phase boundaries identified from numerics. The circular boundary is a conjecture. Only the portion of the triangle where $p_z \geq p_x \geq p_y$ was studied, the rest is obtained by symmetry.}
    \label{fig:layout}
\end{figure*}

Motivated by these developments, in this work we focus on the dynamics and structure of entanglement in a stochastic, measurement-only implementation of Kitaev's honeycomb Hamiltonian. 
We uncover a rich phase diagram, comprised of area-law and critically-entangled phases, both of which have topological qubits, though with different lifetimes. 
Remarkably, in the area-law phases, we find {\it exponentially long-lived} topological qubits. 
This is surprising since the operators being measured do {\it not} form a quantum error-correcting code (in other words, the associated subsystem code is trivial, with zero logical qubits). Moreover, unlike the Floquet code prescription, our dynamics does not follow any specific time-periodic schedule. Nonetheless, it manages to {\it stochastically} generate dynamical qubits that are stable for long times with high probability. 
This phenomenon may be interpreted as a randomized generalization of Floquet codes, and as such may have interesting consequences for developing new approaches to protect quantum information.

The rest of the phase diagram is occupied by a critical phase, which still features topological qubits, albeit with a qualitatively shorter lifetime (polynomial in system size). This phase has several interesting characteristics.
First of all, it feaures a \emph{multiplicative} logarithmic violation of the area-law for the entanglement of late-time states, which sharply distinguishes it from the gapless phase of the static Kitaev Hamiltonian (that has no such violation).
Secondly, the dynamical purification of mixed states~\cite{gullans_dynamical_2020} in this phase obeys a critical scaling with dynamical exponent $z = 1/2$. 
Intriguingly, this represents a \emph{super-ballistic} spreading of information in the dynamics, which is made possible by the non-unitarity and non-locality of measurements.

The rest of the paper is organized as follows.
In Section~\ref{sec:model} we review the Kitaev Hamiltonian and define the measurement-only implementation studied in this work.
With a combination of analytical arguments and numerical simulations based on the stabilizer formalism (reviewed in Appendix~\ref{app:stabilizer_codes}), in Section~\ref{sec:phasediagram} we study the phase diagram of the model, while in Section~\ref{sec:purification} we investigate the nature of each phase through the point of view of dynamical purification. Finally, we summarize our findings and outline directions for future research in Sec.~\ref{sec:conclusion}.


\section{Model \label{sec:model}}

\subsection{ Review of the Kitaev Hamiltonian} \label{subsec:kitaev}
As background, we begin by briefly reviewing the Kitaev honeycomb model~\cite{kitaev_anyons_2006} (readers already familiar with the model can safely skip to Sec.~\ref{subsec:check_structure}). 
The Hamiltonian is given by:
\begin{align} \label{kitaev_hamiltonian}
    \mathcal{H} = -J_x \sum_{x\text{-links}} \sigma_i^x \sigma_j^x - J_y \sum_{y\text{-links}} \sigma_i^y \sigma_j^y - J_z \sum_{z \text{-links}} \sigma_i^z \sigma_j^z.
\end{align}
The spin-$\frac{1}{2}$ degrees of freedom live on vertices of a honeycomb lattice and the interaction terms couple nearest neighbors along the three possible directions, dubbed $\alpha$-links ($\alpha = x, y, z$), as shown in Fig.~\ref{fig:layout}(a).

The ground state of the model can be found exactly by using the presence of conserved $\mathbb{Z}_2$ fluxes at each hexagonal plaquette to reduce the Hamiltonian to free fermions.
When $J_x \approx J_y \approx J_z$, the system is in a gapless spin liquid phase. In this phase, when a magnetic field in the $\hat{x} +\hat{y} + \hat{z}$ direction is applied, the spectrum acquires a gap with non-Abelian anyonic excitations. In contrast, when one coupling is sufficiently larger than the other two, the system is in a gapped phase with Abelian topological order, equivalent to the toric code~\cite{kitaev_anyons_2006}. 

This latter statement can be shown by going deep into the gapped phase, taking (say) $J_z \gg J_x, J_y$ and treating $J_{x,y}$ perturbatively. One can show that in this limit, the Hamiltonian reduces to the toric code~\cite{kitaev_anyons_2006, kitaev_fault-tolerant_2003}. 
Namely, at the exact corner of parameter space where $J_x = J_y = 0$, each $z$-link $(i,j)$ is in an eigenstate of $\sigma_i^z \sigma_j^z = +1$, and one can represent the residual two-dimensional Hilbert space by a single effective qubit, e.g. with Pauli matrices $\tau^z = \sigma^z_i$, $\tau^x = \sigma^x_i \sigma^x_j$. These $\tau$ spins belong on $z$-links of the original honeycomb lattice, which form a square lattice. When $J_x=J_y=0$, any state of the $\tau$ spins is a ground state, giving a massively degenerate ground space. Upon turning on the $J_x, J_y$ couplings, this degeneracy is split at fourth order in perturbation theory, where the following effective Hamiltonian is generated~\cite{kitaev_anyons_2006}:
\begin{align} 
    \mathcal{H}_{\text{eff}} & = -\frac{J_x^2 J_y^2}{16|J_z|^3} \sum_p Q_p \;.
    \label{eq:kitaev_pt}
\end{align}
Here $p$ labels plaquettes and $Q_p$ is the ``plaquette flux'' operator, 
\begin{equation}
Q_p = 
\begin{tikzpicture}[scale=0.5, baseline={([yshift=-.5ex]current bounding box.center)}]
\draw (0,0) -- (0,1) -- (0.866,1.5) -- (1.732,1) -- (1.732,0) -- (0.866,-0.5) -- (0,0);
\draw[anchor=south] (0.866,1.5) node {$\sigma^z$};
\draw[anchor=west] (1.732,1.) node {$\sigma^x$};
\draw[anchor=west] (1.732,0) node {$\sigma^y$};
\draw[anchor=north] (0.866,-0.5) node {$\sigma^z$};
\draw[anchor=east] (0,0) node {$\sigma^x$};
\draw[anchor=east] (0,1) node {$\sigma^y$};
\draw (0.866,0.5) node {$p$};
\end{tikzpicture}
= \begin{tikzpicture}[scale=0.5, baseline={([yshift=-.5ex]current bounding box.center)}]
\draw (0,0) -- (1,1) -- (2,0) -- (1,-1) -- (0,0);
\draw[anchor=east] (0,0) node {$\tau^y$};
\draw[anchor=south] (1,1) node {$\tau^z$};
\draw[anchor=west] (2,0) node {$\tau^y$};
\draw[anchor=north] (1,-1) node {$\tau^z$};
\draw (1,0) node {$p$};
\end{tikzpicture} \;.
\label{eq:plaquette_sketch}
\end{equation}
Here we have rewritten $Q_p$ in the $\tau$ variables by collapsing the two $\sigma$ spins on each $z$-link into a single $\tau$ spin, and turned the plaquette $p$ from a hexagon to a square.
Thus every plaquette of the effective square lattice is assigned a commuting 4-body interaction. Up to single-qubit rotations\footnote{We note that this discussion applies to $L$ even. In this case, the effective square lattice is bipartite, and the stabilizers reduce to the toric code ones after applying suitable single-qubit gates to one sublattice. If $L$ is odd they do not, and there a single logical qubit. Throughout the paper, including all numerical simulations, we have focused on $L$ even.}, this is equivalent to the toric code Hamiltonian~\cite{kitaev_anyons_2006, kitaev_fault-tolerant_2003}.

Before moving on, we remark on an interesting fact concerning the entanglement entropy properties of the Kitaev Hamiltonian. In \textit{both} the gapped and the gapless phases of the model, the contributions to the entanglement entropy can be broken down as contributions due to the ``flux sector'' (the plaquette fluxes and the fluxes through the noncontractible loops in the torus) and contributions due to the remaining degrees of freedom, which can be viewed as fermions~\cite{yao_entanglement_2010}. 
The total entanglement entropy in all phases, measured in bits, is given by 
\begin{align}
    S = \alpha \ell + S_\text{topo},
    \qquad S_\text{topo} = - 1, \label{eq:kitaev_gs_entropy}
\end{align}
where $\alpha$ is a positive constant, $\ell$ is the linear size of the subsystem, and we are dropping terms that vanish at large $\ell$.
Notably, the entanglement entropy follows an area law, and even though the two phases are distinct, 
the topological entanglement entropy is $S_{\text{topo}} = -1$ in both of them\footnote{The topological term comes from the flux sector of the model, which is unaffected by the fermions becoming gapless~\cite{yao_entanglement_2010}.}.

\subsection{Structure of checks} \label{subsec:check_structure}

Unlike the operators which define the toric code Hamiltonian, the operators of the Kitaev honeycomb do not all commute. This precludes treating the Kitaev honeycomb as a stabilizer code. One may instead attempt to view the operators in the Kitaev honeycomb Hamiltonian as ``checks'' in a subsystem code~\cite{poulin_stabilizer_2005, bacon_operator_2006, aliferis_subsystem_2007}. Such codes, reviewed in Appendix~\ref{app:subsystem_codes}, are a generalization of stabilizer codes, quantum codes where a code space is defined to be an eigenstate with eigenvalue $+1$ of a set of operators called stabilizers. In a stabilizer code, error correction proceeds by measuring all of these stabilizers, determining a syndrome by noting which stabilizers yielded a $-1$ value upon measurement, and applying an appropriate error correcting operator. On the other had, subsystem codes differ wherein the stabilizers are not measured directly, but rather in a composite way, as a sequence of measurements of several smaller check operators (which need not commute themselves). A Pauli string is an operator drawn from the Pauli group of $N$ qubits $\mathcal{P}_N$, i.e. it is any product of Pauli operators on independent qubits. Any set of Pauli strings defines a (possibly trivial) subsystem code, and induces a factorization of the system's Hilbert space into {\it stabilizer} qubits, {\it gauge} qubits, and {\it logical} qubits.
One may carry out this construction with the terms in the Kitaev Hamiltonian (Eq. \ref{kitaev_hamiltonian}), i.e. build the subsystem code from check operators, $\mathcal{C} \equiv \{ \sigma_i^\alpha \sigma_j^\alpha \}$ (with $\alpha = x,y,z$ and $(i,j)$ the endpoints of an $\alpha$-link). However, it turns out that this code does not host any logical qubits.

Remarkably, the recent proposal of Floquet codes~\cite{hastings_dynamically_2021} gets around this problem by generating qubits \emph{dynamically}, via measurements of checks in a particular time-periodic sequence.
The implementation of Kitaev's model as a Floquet code is based on a three-coloring of the lattice plaquettes, which partitions the checks into three internally-commuting sets $\mathcal{C}_i$, $i = 0,1,2$; at time $t$, one simultaneously measures all the checks in $\mathcal{C}_{t \text{ mod } 3}$.

In this work we consider a different partition of checks, namely the sets $\mathcal{C}_\alpha = \{\sigma^\alpha_i \sigma^\alpha_j:\ (i,j) \text{ is } \alpha \text{-link}\}$, with $\alpha = x, y, z$, i.e. the three types of terms in Eq.~\eqref{kitaev_hamiltonian}. 
Each element of $\mathcal{C}_\alpha$ commutes with all elements of $\mathcal{C}_\alpha$, and anticommutes with exactly two elements of $\mathcal{C}_\beta$ if $\beta \neq \alpha$. 
This algebra is usefully described in terms of the \emph{frustration graph}~\cite{planat_pauli_2007, chapman_characterization_2020, elman_free_2021, chapman_free-fermion_2022}--a graph whose vertices are Pauli string operators, and whose edges connect any two anticommuting operators. The frustration graph of $\mathcal{C}$ is a kagome lattice, with each $\mathcal{C}_\alpha$ forming a triangular sublattice (Fig.~\ref{fig:layout}(b)). Notably, the solvability of Kitaev's Hamiltonian in terms of free fermions can be derived purely from properties of this graph (namely from the fact that it is a ``\emph{line graph}'')~\cite{chapman_characterization_2020}.

As the frustration graph picture shows, any product of checks along a closed loop on the lattice (whether contractible or not) is a \emph{stabilizer}, that is to say, is a product of checks that commutes with all checks. In particular the product of checks around any hexagonal plaquette is a stabilizer\footnote{Such stabilizers are precisely the $Q_p$ plaquette fluxed employed to solve Kitaev's Hamiltonian. (Eq.~\eqref{kitaev_hamiltonian})} and so is the product of checks along a nontrivial loop on the torus\footnote{In the Floquet code implementation such topologically nontrivial stabilizers are never measured, and thus can be used to store a logical qubit.}.
Conversely, products of checks along open strings anticommute with four checks (two near each endpoint). Note that products of checks around triangles in the frustration graph are \textit{not} loops. These contain three open endpoints.
This structure will be useful for the following analysis of measurement-only dynamics.

\subsection{Measurement-only dynamics} \label{subsec:mom_dynamics}

We study the measurement-only dynamics~\cite{ippoliti_entanglement_2021} induced by random measurements of the check operators in $\mathcal{C}$, as discussed above.
We focus on honeycomb lattices of $L\, \times \, L$ unit cells, each one comprising two qubits, giving $N = 2L^2$ physical qubits (Fig \ref{fig:layout}(a)). 
The system is initialized either in a product state $\ket{0}^{\otimes N}$ (that is, each spin $s_i$ is initialized in the $+1$ eigenstate of the local $S^i_z$ operator), or in a fully-mixed state, in order to study entanglement phases and dynamical purification phases respectively. 
At each time step, one of $\mathcal{C}_x$, $\mathcal{C}_y$, $\mathcal{C}_z$ is chosen with probability $p_x, p_y$ and $p_z$ respectively. Once a set $\alpha$ is chosen, a single check from $\mathcal{C}_\alpha$ is chosen uniformly at random. 
We define a unit of time to consist of $N$ measurements, such that each check is measured once on average per unit time. We evolve the initial state for an amount of time to be specified in each case, but always $\text{poly}(N)$. 

The parameter space of our measurement-only dynamics is defined by probabilities $p_x, p_y$ and $p_z$. 
In this parameter space, the pure-state dynamics realizes a rich phase diagram including a topological critical phase, topological area-law entangled phases, as well as area-law phases and critical points where the system reduces to decoupled one-dimensional wires; each of these can also be analyzed as a dynamical purification phase.
In the next section we explore this phase diagram with a combination of analytical arguments based on the structure of checks, and of numerical simulations based on the stabilizer method. This is an efficient computational approach enabled by the fact that all the 2-qubit measurements involved in this circuit are Pauli strings; a brief review is provided in Appendix~\ref{app:stabilizer_codes}.


\section{Phase Diagram \label{sec:phasediagram}}

In order to determine the phase diagram of the model, we look at two diagnostics, both starting from pure initial product states: (i) the scaling of half-system entanglement entropy with system size at late times, and (ii) the saturation value of the tripartite mutual information for a tripartite subsystem of nontrivial topology.
The former diagnostic distinguishes an area-law phase ($S \sim L$) from a volume-law phase ($S \sim L^2$) and potentially from a critical point, where logarithmic violations of the area-law ($S \sim L\ln(L)$) have been reported~\cite{nahum_entanglement_2020, lavasani_topological_2021, turkeshi_measurement-induced_2020}.
The latter, in the geometry shown in \ref{fig:I3}(a), similarly distinguishes area-law, volume-law and critical states, but additionally detects the topological entanglement entropy~\cite{kitaev_topological_2006}: $\mathcal{I}_3 = 2S_\text{topo} + \delta \mathcal{I}_3$, where $S_\text{topo}$ is the universal additive term to the entropy as in Eq.~\eqref{eq:kitaev_gs_entropy}.
$\delta \mathcal{I}_3$ is the non-topological contribution, which vanishes in area-law states and is extensive in volume-law states. Its behavior in critical states is more subtle and is discussed in Appendix~\ref{app:I3}.

Before discussing these numerical diagnostics, we begin by noting that, along the three sides of the phase diagram (defined by $p_\alpha = 0$ for $\alpha = x, y, z$), only two types of checks are ever measured; thus the model effectively splits into decoupled one-dimensional wires whose dynamics can be solved analytically. 
As can be seen from the frustration graph (Fig.~\ref{fig:layout}(b)), each wire's monitored dynamics is equivalent to that of a monitored Ising model~\cite{nahum_entanglement_2020, lavasani_measurement-induced_2021, ippoliti_entanglement_2021, sang_measurement_2020, lang_entanglement_2020}. Thus the sides of the phase diagram are in an area-law phase, except their mid-points (e.g. $p_x = p_y = 1/2$, $p_z=0$) which are critical, and described by loop percolation. 

In the following we focus on the interior of the phase diagram. A summary of our results is shown in Fig.~\ref{fig:layout}(c).
We focus our discussion on the line cut in parameter space parametrized by $0 \leq p_z \leq 1$, with $p_x = p_y = \frac{1}{2}(1-p_z)$; phase boundaries along different directions are obtained analogously. 

\subsection{Entanglement entropy \label{sec:ee}}

\begin{figure} 
    \centering
    \includegraphics[trim={1cm 0 0 0},width = \linewidth]{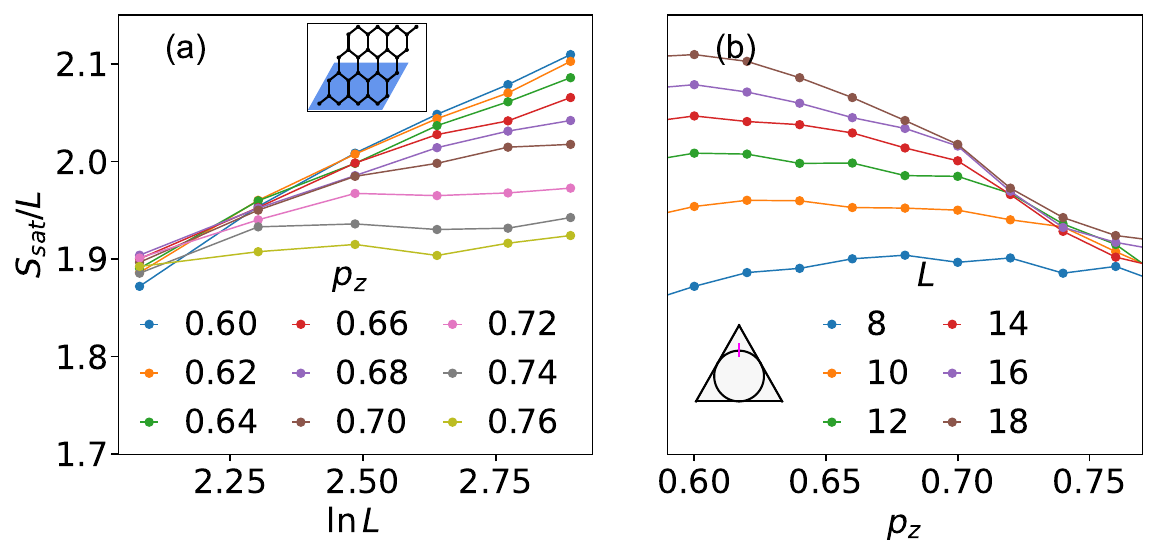}
    \caption{ Mapping the phase diagram with the half-system entanglement entropy.
    (a) Saturation value of the half-system entropy at late times, $S_\text{sat}$, normalized by linear system size $L$, at different points in the phase diagram $p_x = p_y = (1-p_z)/2$. 
    Saturation to a constant indicates area-law behavior, while a finite slope (vs $\ln(L)$) indicates a logarithmic violation of the area-law.
    The data is averaged over 100 realizations. The longest time simulated is $t = 50N$. This limits the exploration of larger $p_z$, since the dynamics severely slow down as $p_z = 1$ is approached.
    Inset: geometric layout. The shaded region denotes the subsystem. 
    (b) Same data shown as a function of $p_z$, with different curves representing different sytem sizes. Overlap between curves for different sizes corresponds to an area-law.
    Inset: parameter space. The pink line highlights points studied in this figure.     }
    \label{fig:phase_analysis}
\end{figure}

The scaling of the saturation value of the entanglement entropy, $S_\text{sat}$, divided by the length of the system $L$, for a subsystem $\{0\leq x < L/2\}$ consisting of half the system, is shown in \ref{fig:phase_analysis}(a).
For $p_z \lesssim 0.68$, we observe an approximately linear relationship between $S_{\text{sat}}(L)/L$ and $\ln{L}$, a logarithmic violation of the area-law indicative of a critical phase, but unlike the corresponding Hamiltonian phase (where the entanglement remains area-law). 
For $p_z \gtrsim 0.68$, the $S_\text{sat}(L)$ curves begin to exhibit saturation in $L$, indicating area law behavior. 
This is further highlighted in Fig.~\ref{fig:phase_analysis}(b), where we show the normalized values $S_\text{sat}(L)/L$ as a function of $p_z$. At $p_z > 0.68$, the curves for the different system sizes begin to collapse onto each other, indicating a transition into the area law phase. We cannot capture the full scaling behavior close to $p_z = 1$ because the time required for the entanglement entropy to saturate becomes very long (in fact this is apparent already for $p_z \gtrsim 0.72$ and accounts for the drift between the system sizes at $p_z = 0.74,0.76$). This signals a very slow dynamics in the area law phase, to which we shall return in Sec.~\ref{sec:purification}. 
Although from this analysis we can estimate the location of the phase boundary at some $p_z$ in the $0.65-0.70$ range, we note that this diagnostic is particularly susceptible to finite size effects.
This motivates considering a different diagnostic.

\subsection{Tripartite mutual information \label{sec:i3}}

\begin{figure*} 
    \centering
    \includegraphics[width=\textwidth]{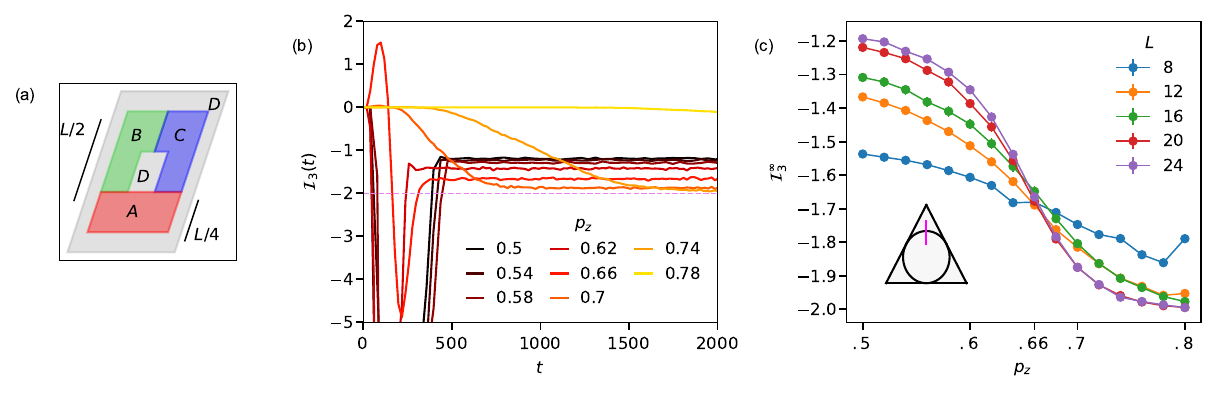}
    \caption{Mapping the phase diagram with the tripartite mutual information $\mathcal{I}_3$. 
    (a) Geometry of the subsystems used to compute $\mathcal{I}_3$. $A$, $B$, and $C$ are red, green and blue respectively. 
    (b) Time dependence of $\mathcal{I}_3$ starting from a disentangled pure state $\ket{0}^{\otimes N}$, shown for a system size of $L = 24$, averaged over $10^2-10^3$ realizations depending upon $p_z$. For larger $p_z$, greater maximum evolution times were required to reach saturation.  
    In the critical phase, the steady-state plateau is offset away from the topological value of $-2$. 
    (c) The plateau value as a function of $p_z$ for multiple system sizes. As $L$ increases, this transition seemingly sharpens, with the crossing between curves indicating the phase boundary. The time to reach steady state varies depending on system size and $p_z$; we simulated a maximum evolution time of $t = 20N$. Each point here is the average of $10^2-10^5$ realizations depending upon system size.
    Inset: parameter space, with approximate phase boundary (circle) and the parameter interval studied (pink segment).
    } 
    \label{fig:I3}
\end{figure*}

As another diagnostic of the phases and phase boundaries, we use the tripartite mutual information, which has been introduced as a means of extracting the topological entanglement entropy in two-dimensional ground states~\cite{kitaev_topological_2006}. 
We partition the system into four regions $A$, $B$, $C$ and their complement, as shown in the inset to Fig.~\ref{fig:I3}(a), and calculate the following:
\begin{align}\label{eq:mutual info}
    \mathcal{I}_3 = S_A + S_B + S_C - S_{AB} - S_{BC} - S_{AC} + S_{ABC}.
\end{align}
This quantity has been successfully employed in the study of measurement-induced phase transitions in one spatial dimension~\cite{gullans_dynamical_2020, zabalo_critical_2020, ippoliti_entanglement_2021} due to the absence of logarithmic divergence at the critical point, which limits finite-size drifts. We note that the geometry we chose, where $ABC$ is topologically nontrivial, causes the topological entanglement entropy $S_\text{topo}$ to appear twice in $S_{ABC}$ (as the complementary subsystem has two connected components); thus $S_\text{topo}$ appears 5 times with a positive sign and 3 times with a negative sign, giving $\mathcal{I}_3 = 2S_\text{topo} + \delta\mathcal{I}_3$, as mentioned earlier. For a state with toric code topological order ($S_\text{topo}=-1$), we expect $\mathcal{I}_3 = -2 + \delta\mathcal{I}_3$.

Fig \ref{fig:I3}(a) shows the time dependence of $\mathcal{I}_3$ (the initial state $\ket{0}^{\otimes N}$ has $\mathcal{I}_3 = 0$). 
We see that for all values of $p_z$, $\mathcal{I}_3$ eventually achieves a steady value.
In the critical phase, the time scale for convergence to the plateau is relatively short, but as the system gets deeper into the area law phase the time scale rapidly increases.
Fig.~\ref{fig:I3}(b) shows the value of the late-time plateau in $\mathcal{I}_3$ (collected after having evolved the system to saturation) as a function of $p_z$. As $p_z$ is varied from 0.5 to 0.8, there is a smooth crossover from the predicted topological value of $-2$ in the area law phase, to a different and significantly higher value. 
In Appendix~\ref{app:I3} we show that the critical contribution $\delta \mathcal{I}_3$ approaches a finite non-zero value in the critical phase. Thus this crossover is indicative of the area-law to critical phase transition.

We estimate the location of the phase boundary based on the crossing point between curves for different system sizes, e.g. $p_z = 0.66(1)$ for the curves shown in Fig.~\ref{fig:I3}(c) ($p_x = p_y = 1-p_z$).  
Performing a similar analysis for other line cuts in parameter space yields the phase diagram shown in \ref{fig:layout}(c).

\begin{figure}
    \centering
    \includegraphics[]{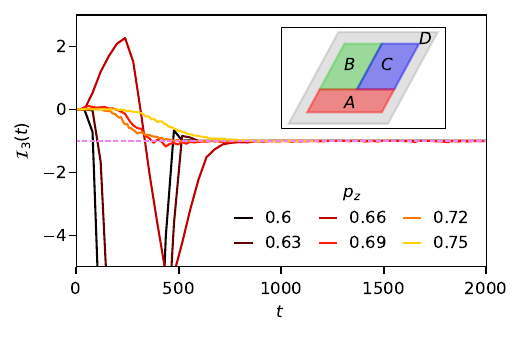}
    \caption{Tripartite mutual information as a function of time, starting from a disentangled pure state $\ket{0}^{\otimes N}$, for a choice of subsystems $A$, $B$, $C$ such that $ABC$ is topologically trivial (shown in inset). The system size is $L=24$ and each curve is the average of $10^2$ realizations. This subsystem geometry isolates only the topological contribution, and thus $\mathcal{I}_3$ asymptotes to the same value ($S_\text{topo} = -1$) in both phases.}
    \label{fig:I3_topo}
\end{figure}

As noted above, $\mathcal{I}_3$ picks up two types of contributions: topological and critical. The latter (which account for the positive deviation from $-2$ in the critical phase) arise due to the fact that with the geometry employed in Fig.~\ref{fig:I3}, the complement to $ABC$ is disconnected. This introduces both a critical contribution, and a double-counting of the topological entanglement entropy $S_\text{topo}$, as we discuss in Appendix~\ref{app:I3}.
In Fig.~\ref{fig:I3_topo} we instead consider a geometry where $ABC$ is topologically trivial, such that there is no critical contribution, and $S_{\text{topo}}$ appears only once on net. 
Thus we see that in both the critical ($p_z < 0.66$) and the area-law phases ($p_z > 0.66$), $\mathcal{I}_3$ plateaus at exactly $S_\text{topo} = -1$. This is explained as follows: In both the critical and area law phases, stabilizers (i.e. products of checks along closed loops, both topological and trivial, making up the ``flux'' sector of the model) are randomly generated over the course of the dynamics. Once these stabilizers are generated, they cannot be destroyed as they commute with all the check measurements. The topological qubit is encoded in the Wilson loops of the model, that is in the degrees of freedom corresponding to the product of checks on a \textit{homologically non-trivial loop} (a loop which wraps around the entire system). This loop is generated in all phases of the model, and once generated it cannot be destroyed. In this sense, \emph{both the critical and area-law phases of this model are topological}. 


\section{Dynamics in the Two Phases \label{sec:purification}}

Having shown that both phases possess a topological sector that can, in principle, store quantum information, we now aim to characterize the stability of the topological qubits in each phase over time.
To this end, and to further understand the dynamics in each phase more broadly, we study the problem of dynamical purification of mixed states~\cite{gullans_dynamical_2020}, which is closely related to the problem of entanglement scaling in late-time states. 
A mixed initial state $\rho(0)$, subject to the monitored dynamics, eventually loses its entropy and becomes pure. However, the time taken to purify, $\tau$, defines \textit{purification phases}. If the system purifies quickly, in a time scale $\tau = O(\log(N))$, it is said to be in a {\it pure phase}; conversely, if it retains memory of the initial condition of a long time $\tau = \mathcal{O}(\exp{N})$, it is said to be in a {\it mixed phase}~\cite{fidkowski_how_2021}. These phases usually correspond, respectively, to area-law and volume-law entanglement phases in the dynamics of pure states.
Finally, one has critical points between the aforementioned phases, with an algebraic scaling of the purification time, $\tau \sim L^z$.
The purification of a mixed state corresponds to the loss of any information that may have been initially encoded in it, and is thus directly relevant to the stability of the topological qubits~\cite{gullans_dynamical_2020}.

Here, we focus on stabilizer states, and thus consider initial states defined by an incomplete list of stabilizers: $\rho = \prod_i (\mathbb{I}+s_i)/2$, where the list of generators $\{s_i\}$ has length $N-S$, and $S$ is the state's entropy in bits. In the case of a completely mixed state ($S=N$), the list is empty, and the density matrix takes the form $\rho = \mathbb{I} / 2^N$. 
As the monitored circuit dynamics proceeds, new stabilizer generators are added according to a set of rules reviewed in Appendix~\ref{app:stabilizer_codes}, and the system eventually purifies ($S=0$) when the list of generators achieves maximal length $N$.

\subsection{Critical phase \label{sec:crit_purif}}

\begin{figure}
    \centering
    \includegraphics[width=\columnwidth]{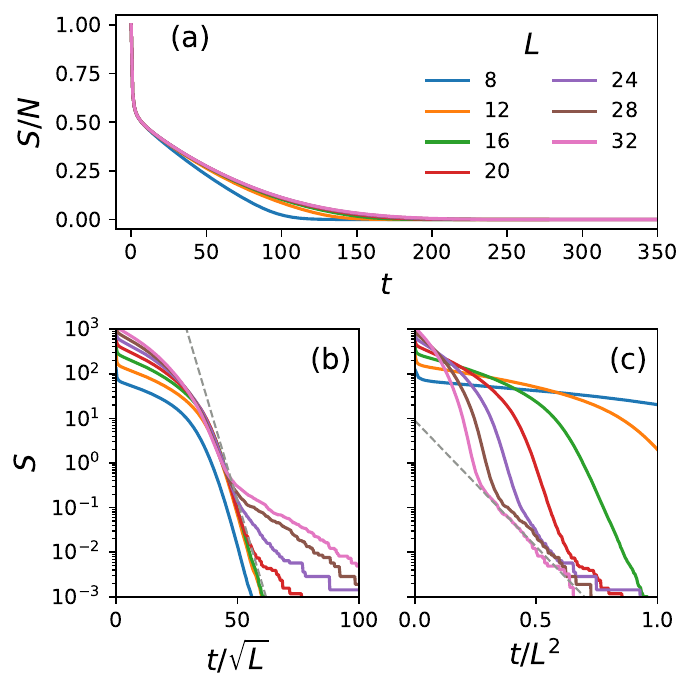}
    \caption{
    Purification dynamics in the critical phase ($p_x = p_y = p_z = 1/3$). 
    Data from stabilizer simulations averaged over between $10^5$ and $10^3$ realizations depending on size. 
    (a) Entropy density $S/N$ as a function of time $t$. 
    (b) Entropy $S$ as a function of rescaled time $t/\sqrt{L}$. The data are consistent with a form $S\sim e^{-c t/\sqrt{L}}$ (dashed line) with the exception of $S<1$ values at the largest sizes, where the purification visibly slows down.
    (c) Entropy $S$ as a function of rescaled time $t/L^2$. At late times and for the largest system sizes, the data become consistent with a form $S\sim e^{-c t / L^2}$ (dashed line).
    }
    \label{fig:purification_dynamics}
\end{figure}

We begin by analyzing the critical phase purification dynamics starting form the completely mixed state.
The results of numerical simulations of the purification dynamics at the center of the critical phase, $p_x = p_y = p_z = 1/3$, are shown in Fig.~\ref{fig:purification_dynamics}.
Fig.~\ref{fig:purification_dynamics}(a) shows that, at early times, the entropy density $S/N$ is system-size independent and drops nearly-immediately from 1 to $\approx 0.5$. Following this initial drop, we see a somewhat slower decay towards zero entropy density, accompanied by a separation between different system sizes.
The intermediate-time behavior does {\it not} show a power-law decay of the entropy in time. This is unlike what is found in measurement-induced critical points or phases that admit a CFT description\footnote{We note that similar behavior was recently seen in a non-conformal critical point in two dimensions with dynamical exponent $z = 3/2$~\cite{lavasani_topological_2021}.}~\cite{gullans_dynamical_2020, li_conformal_2021, ippoliti_entanglement_2021, lavasani_topological_2021}.
However, the late-time decay displays critical scaling.  Fig.~\ref{fig:purification_dynamics}(b)
shows a crossover from the early-time behavior $S(t) \sim L^2 f(t)$ to an intermediate-time scaling behavior\footnote{We note that an exponential decay $S \sim e^{-t/L}$ at late times is expected in critical points described by CFTs~\cite{li_conformal_2021}; exponential decay $S\sim e^{-t/L^z}$ in non-CFT critical phases was also observed~\cite{ippoliti_fractal_2022}.}
\begin{align}
    S(t) \sim e^{-t / \sqrt{L}}.
\end{align}

Intriguingly, this corresponds to a dynamical exponent of $z = 1/2$.
Note that this is {\it not} a diffusive scaling ($t \sim L^2$, $z=2$), but rather a kind of ``superluminal'' one: for instance, we may use the purification of a local probe qubit~\cite{gullans_scalable_2020} to learn about the size $L$ of the system in time $t\sim \sqrt{L}$, parametrically faster than what would be allowed by a causal light cone. However, in order to do so, we implicitly need access to measurement outcomes across the whole system, thus preventing superluminal signaling. 
We also note that non-conformal values of $z<1$ were recently found in long-range-interacting monitored systems~\cite{block_measurement-induced_2022} and in ``space-time duals'' of unitary circuits in a Griffiths regime near a localization transition~\cite{ippoliti_fractal_2022}.
However, in both cases the $z$ exponent varies continuously as a function of model parameters (an interaction range exponent and proximity to the locaization transition, respectively). 
Here instead, the value $z = 1/2$ is a feature of the entire critical phase.

Finally, at the largest system sizes and latest time scales (corresponding to residual entropy below 1 bit), we see a deviation from the aforementioned $z = 1/2$ scaling. This is found to be the onset of a different critical regime, with dynamical exponent $z=2$, as highlighted in Fig.~\ref{fig:purification_dynamics}(c). 
We understand the coexistence of these two critical behaviors as coming from the flux and fermionic sectors of the model, respectively. Namely, if the flux sector becomes completely purified before the fermionic sector does, residual open-string (fermionic) operators undergo critical dynamics with $z=2$. 
This is discussed further in Appendix~\ref{app:purification_times}, where we also analyze the distribution of purification times $P(\tau)$. 

In either case, a scaling $\tau \sim L^z$ implies that any information encoded in the model's topological qubits will be corrupted in a $\text{poly}(L)$ time scale.

\subsection{Area-law phase \label{sec:area_purif}}

\begin{figure}
    \centering
    \includegraphics[width=\linewidth]{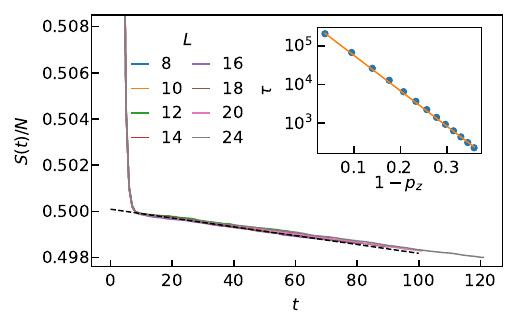}
    \caption{Purification dynamics in area law phase at $p_z = 0.9, p_x = p_y = 0.05$. 
    Data from stabilizer numerical simulations, averaged over $10^3$ to $10^4$ realizations depending upon system size. Maximum evolution time used here was $t = 5L$.
    Main plot: entropy density $S/N$ as a function of time. 
    Inset: purification time $\tau$, extracted from fits to sizes $L \leq 16$, as a function of $p_z$.     }
    \label{fig:purification_dynamics_area}
\end{figure}

We now move on to the area-law phase, with numerical results shown in Fig.~\ref{fig:purification_dynamics_area} for $p_x = p_y = 0.05$, $p_z = 0.90$. 
We again find that the early-time dynamics shows a sharp drop from $S = 2L^2$ to $S \approx L^2$; following this drop, we find that the entropy obeys the scaling
\begin{align}
    S(t) \sim L^2 e^{-t/\tau},
\end{align}
with a size-independent purification time scale $\tau$, as expected in a pure phase~\cite{gullans_dynamical_2020}. 
However, we observe that $\tau$ diverges as we approach the corner of the phase diagram $p_z = 1$; for example, $\tau \approx 2.4\times10^4$ when $p_z = 0.9$. 

These slow dynamics can be understood by starting from $p_z = 1$. At this extreme point, the system does not purify, but instead the entropy density remains constant at $0.5$: since we are only measuring one type of checks ($\mathcal{C}_z$), there is one bit of information for each $z$-bond that is not revealed by the measurements (see also our discussion of this limit in Sec.~\ref{subsec:kitaev}). 
We can then consider how this is modified when $p_x$ and $p_y$ are small but finite. In purification dynamics for stabilizer states, entropy decreases only when a measurement commutes with all the existing stabilizers (while not being generated by them): this adds an independent element to the list of stabilizer generators and thus lowers the entropy by 1. 
When a check in in $\mathcal{C}_x$ or $\mathcal{C}_y$ is measured, it is highly likely to anticommute with some existing stabilizers, as elements of $\mathcal{C}_z$ are frequently measured and likely to be in the stabilizer group; thus the new measurement merely replaces one of the stabilizer generators with which it anti-commutes, conserving the entropy of the state (see also Appendix~\ref{app:stabilizer_codes} for further details on the update rules). This new $x$- or $y$-link stabilizer is then likely to be immediately removed by subsequent $z$-link measurements, reverting to the initial state. Entropy does not change in this process.

However, if a suitable sequence of {\it four} measurements of the infrequent kinds occurs, then it is possible to lower the entropy of the state: namely, by forming a $Q_p$ ``plaquette flux'' stabilizer, as defined in Eq.~\eqref{eq:plaquette_sketch}. 
Once added to the stabilizer group, such operator cannot be removed, and the entropy has permanently decreased by 1. 
This is a fourth-order process, as it requires that four $\mathcal{C}_x/\mathcal{C}_y$ bonds be measured in a row; thus we expect $\tau^{-1} \sim p_x^2 p_y^2 = (1-p_z)^4$. 
This expectation is borne out in numerics, as shown in Fig.~\ref{fig:purification_dynamics_area} (inset).
From fits to the data at various points on the $p_x = p_y$ cut of the phase diagram we find
\begin{align}
    \tau(p_z) \sim (1-p_z)^{-c}, \qquad c = 4.2(1),
\end{align}
in agreement with our analytical argument.

Suppose now one waits long enough so that all the plaquette fluxes $Q_p$ have been added to the stabilizer group. This will occur after $t\approx (1-p_z)^{-4} \ln(L)$.  
At this point, the stabilizer group is generated by all the plaquette fluxes $Q_p$ and by products of checks along short open strings (i.e., $\mathcal{C}_z$ bonds perturbed by a few $\mathcal{C}_{x,y}$ ``errors''). The only stabilizers remaining to be added are the topological ones: products of checks along noncontractible loops on the torus.
In order for such stabilizers to be added, enough $\mathcal{C}_x$ and $\mathcal{C}_y$ measurements must occur to connect around the entire system. With high probability, this requires a time that is exponentially long in $L$. Therefore, a remaining finite amount ($2$ bits) of entropy will fail to purify up to this timescale. 
In this phase, the model is thus a stochastic, dynamical version of a topological quantum memory. 

\begin{figure}
    \includegraphics[width = \linewidth]{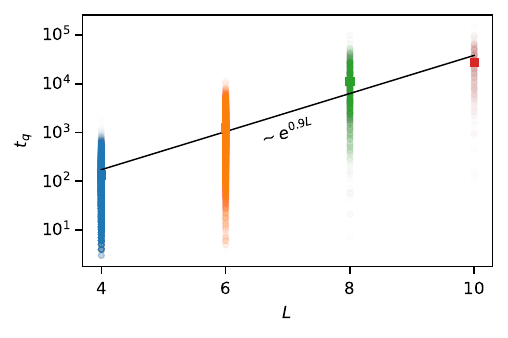}
    \caption{Purification of the topological qubits in the area-law phase. $t_q$ denotes the time elapsed between reaching $S=2$ and $S=1$, i.e., the lifetime of the first topological qubit. Circles represent $t_q$ for individual runs of the dynamics (between $10^2$ and $10^4$ runs are shown depending upon size). Squares represent the average over runs. The line is an exponential fit to the average $t_q$ values.
    }
    \label{fig:weakly_mixed}
\end{figure}

In order to resolve this effect numerically, the slow purification of plaquette fluxes ($\tau \sim (1-p_z)^{-4}$) is a hindrance. We thus consider an initial mixed state where all the plaquette fluxes have been measured. The entropy of such a state quickly converges to $S = 2$, and subsequently remains at that value for a time $t_q$ which scales exponentially in $L$, as shown in Fig.~\ref{fig:weakly_mixed}. 
This same exponential time scale causes the slow-down of the convergence of $\mathcal{I}_3(t)$ to its asymptotic value ($2S_\text{topo}$ or $S_\text{topo}$, respectively) previously observed in Figs.~\ref{fig:I3} and \ref{fig:I3_topo}, upon entering the area-law phase. The purification of topological qubits in the mixed-state dynamics and the generation of topological entanglement entropy in the pure-state dynamics are both described by the same process (an uninterrupted sequence of ``errors'' that wrap around the torus), whose characteristic time scale diverges exponentially in system size.

\subsection{Relationship with Hamiltonian phases}

We now remark on similarities of our dynamical measurement circuit model with that of the static Hamiltonian. 
In the area-law phase, 
where one kind of measurement is dominant over the others, e.g. $p_z \gg p_{x,y}$, we find that the formation of the plaquette stabilizers occurs over a time scanel {\it quartic} in $p_{x,y}$.
This closely mirrors the Hamiltonian solution, where the plaquette operators also appear at fourth order in perturbation theory, Eq.~\eqref{eq:kitaev_pt}. The reason for this quartic scaling is the same: commutation with the dominant bonds, which in the Hamiltonian perturbation theory arises from the ``on-shell'' requirement for the energy of the process while in the monitored dynamics arises from the requirements for increasing the number of stabilizers.
Once the plaquette stabilizers are formed, 
the system is described by a toric code in the common eigenspace of all $\mathcal{C}_z$ checks. 
Thus the topological entanglement entropy of both the measurement circuit and the Kitaev Hamiltonian have the same origin. 

The critical phase, on the other hand, is sharply different between the monitored and Hamiltonian models. This is immediately visible from the logarithmic violation of the area-law, present in the former but not the latter.
While the flux sector (i.e., the plaquette and topological stabilizers) is identical in the two cases, the state of the remaining degrees of freedom is different. The Hamiltonian ground state can be written in terms of free fermions in a Dirac band structure at charge neutrality. The monitored steady states instead feature open strings (which can be thought of as two-Majorana operators) that are localized in real space. The logarithmic violation of the area-law comes from their long-ranged length distribution (which can be inferred from the critical term in $\mathcal{I}_3$, see Appendix~\ref{app:I3}), not from the presence of a Fermi surface. 
Finally, we note that the shape of the phase boundary changes, from triangular in the Hamiltonian model~\cite{kitaev_anyons_2006} to seemingly circular in the measurement-only model, Fig.~\ref{fig:layout}(c). Given the simplicity of the model, a regular shape for the phase boundary appears plausible; it would be interesting to prove that the shape is indeed circular in this case.


\section{Conclusion and Outlook \label{sec:conclusion}}

To summarize, in this work we have studied a class of measurement-only circuit dynamics based on Kitaev's honeycomb Hamiltonian and analyzed its phase structure based on the entanglement of pure states and the purification of mixed states. We found a phase diagram consisting of two distinct kind of phases: area law and critical, the latter exhibiting an entanglement scaling of the form $S_A \sim \ell \log{\ell}$ for a subsystem $A$ of linear size $\ell$. Both phases are topological, as measured for example by their topological entanglement entropy of $S_\text{topo} = -1$, which indicates the presence of topological qubits.

In the area law phase, we found that the lifetime of this qubit scales exponentially with system size, making these stochastic analogues of Floquet codes~\cite{hastings_dynamically_2021}, 
while in the critical phase we found a lifetime that is only polynomial.
The same exponential lifetime was recently observed in a measurement-only version of a perturbed toric code~\cite{lavasani_topological_2021}. However, the honeycomb model studied here lacks an explicit commuting-projector limit, and thus even in the absence of errors or perturbations, there is nontrivial dynamics taking place, with incompatible (i.e., noncommuting) measurements being repeated indefinitely and preventing relaxation to a steady state. This makes the emergence of protected logical qubits more surprising.

A distinctive feature of the critical phase arises in the dynamical purification of mixed states: we found that, after an initial regime where the entropy density is system size independent (na\"ively indicative of a {\it pure} phase~\cite{gullans_dynamical_2020}), the late-time scaling behavior yields a dynamical exponent $z = 1/2$.
This is remarkable as a value of $z<1$ implies a ``superluminal'' scaling, which is normally not allowed on grounds of locality (e.g. by Lieb-Robinson bounds~\cite{lieb_finite_1972}). Here it becomes possible due to the presence of measurements, whose effects are nonlocal. We note however that quantum measurements are non-signaling, so the need for classical communication of measurement outcomes across the whole system ultimately restores causality. The algebraic scaling $t \sim L^z$ also governs the lifetime of the topological qubit in this phase. Thus, in this case, their information content is corrupted by the monitored dynamics over a time $\text{poly}(L)$.

The area law phase we found bears a strong resemblance to the recent idea of Floquet codes~\cite{hastings_dynamically_2021}. 
There, logical qubits are generated dynamically by a periodically-repeated sequence of measurements. 
It is interesting to note that this stability largely survives in the stochastic setting studied in this work (with the qubit lifetime being exponential in system size rather than infinite).
This raises several interesting questions. 
A key aspect of the Floquet code proposal is the existence of a fault-tolerance threshold, which can be derived with the same statistical-mechanics mapping used in the time-independent case~\cite{dennis_topological_2002}. Is it possible to devise a fault-tolerant decoding scheme for this {\it stochastic} quantum memory? If so, could the randomness enhance performance in certain cases (e.g. against biased noise, as seen in other randomized constructions recently~\cite{dua_clifford-deformed_2022})?
Conversely, what other phases of monitored circuits may be ``de-randomized'' to yield useful quantum error-correcting codes?

From a perspective more familiar to a condensed matter physicist, our results raise questions about the structure of phases in monitored circuits. Should one think of the area-law phases of our stochastic model, the measurement-only toric code of Ref.~\cite{lavasani_topological_2021}, and the Floquet code of Ref.~\cite{hastings_dynamically_2021} as belonging to the same phase of matter? 
In unitary dynamics, time periodicity can give rise to novel phases with no static analogues~\cite{rudner_anomalous_2013, khemani_phase_2016, harper_topology_2020}.
What is the role of time periodicity in monitored dynamics? In particular, can time-periodic measurement-protocols realize distinct phases that do not have stochastic realizations? 

Our results point to several other directions for future research. 
Kitaev's honeycomb Hamiltonian is a celebrated exactly-solvable model of topological condensed matter physics, with a rich and interesting structure. The monitored circuits studied in this work reflect this structure, while also exhibiting some distinct and novel aspects. 
This raises the question of whether other paradigmatic models in condensed matter physics may admit similarly interesting non-equilibrium versions, and what novel phenomena these may give rise to. 
Thus far, models that have been analyzed through this lens include the 1D Ising model~\cite{nahum_entanglement_2020, lavasani_measurement-induced_2021, ippoliti_entanglement_2021, sang_measurement_2020, lang_entanglement_2020}, the 2D toric code~\cite{lavasani_topological_2021}, and the honeycomb model in this work. 
Models in three or more spatial dimensions remain largely unexplored. Notably these include lattice Hamiltonian models of {\it fractons}~\cite{vijay_new_2015, nandkishore_fractons_2019}, whose slow dynamics in the Hamiltonian setting may translate to monitored circuits with distinctive behavior.
Furthermore, all models listed above are built out of spin-$1/2$ degrees of freedom, or qubits; it would be interesting to explore monitored versions of models with higher-dimensional qudits (including e.g. the spin-1 AKLT chain~\cite{affleck_rigorous_1987}) or bosonic degrees of freedom.

{\it Note added.} During completion of this manuscript, a related work studying the same model appeared~\cite{lavasani_monitored_2022}.

\begin{acknowledgments}
This work was supported with funding from the US Department of Energy, Office of Science, Basic Energy Sciences, under Early Career Award No. DE-SC0021111 (A. S. and V.K.). V.K. also acknowledges support from the Sloan Foundation through a Sloan Research Fellowship and the Packard Foundation through a Packard Fellowship. M.I. was funded in part by the Gordon and Betty Moore Foundation's EPiQS Initiative through Grant GBMF8686.  T.R. is supported in part by the Stanford Q-Farm Bloch Postdoctoral Fellowship in Quantum Science and Engineering.  Numerical simulations were performed on Stanford Research Computing Center's Sherlock cluster.
\end{acknowledgments}

\appendix
\section{Stabilizer Simulations \label{app:stabilizer_codes}}

Here we review the simulation of stabilizer states under Clifford circuits. Recall that a stabilizer state is an $N$-qubit state that is the $+1$ eigenstate of operators belonging to an Abelian subgroup $\mathcal{S}$ of the $N$-qubit Pauli group $\mathcal{P}_N$:
\begin{align}
    S\ket{\psi} = \ket{\psi}, \forall S \in \mathcal{S}.
\end{align}
This group $\mathcal{S}$ has $2^N$ elements, but is minimally generated by a set of $N$ independent elements, $\{g_i\} \subset \mathcal{P}_N$.

Stabilizer states are useful because of their compact classical representation (they are specified by $N$ Pauli strings, i.e. $O(N^2)$ classical bits, unlike generic states whose representation requires $\exp(N)$ bits), while also capturing many important properties of quantum information such as entanglement.
Unitary transformations that preserve stabilizer states form the Clifford group~\cite{aaronson_improved_2004}, which may be generated by the Hadamard, phase and CNOT gates:
\begin{align}
    H &= \frac{1}{\sqrt{2}} \begin{pmatrix}1&1\\1&-1\end{pmatrix}, \\
    P &= \begin{pmatrix}e^{-i\pi/4} & 0 \\ 0 & e^{i\pi/4} \end{pmatrix}, \\
    CNOT &= \begin{pmatrix} 1 & 0 & 0 & 0 \\0 & 1 & 0 & 0 \\ 0 & 0 & 0 & 1 \\ 0 & 0 & 1 & 0 \end{pmatrix}.
\end{align}
By the Gottesman-Knill theorem, stabilizer circuits (circuits with gates sampled only from the Clifford group) are efficiently simulable on classical computers in a time that scales as $\mathcal{O}((d+n)N^2)$, where $d$ is the circuit depth and $n$ is the number of measurements per layer \cite{aaronson_improved_2004}. 
Moreover, Clifford group elements have an efficient representation over $GF(2)$, and as such, there exists a relatively simple algorithm, involving only linear algebra over $\mathds{Z}_2$, for keeping track of the list of stabilizers upon evolution by each gate. This algorithm and the update rules are detailed in \cite{aaronson_improved_2004}. 
Here we review the update rules that are most relevant to understand measurement-based circuits (also reviewed e.g. in Ref.~\cite{gullans_dynamical_2020, ippoliti_entanglement_2021}). 
Suppose we have a stabilizer state, with stabilizer generators given by $\{ g_1,g_2...g_N \}$, and we measure an operator $O$. If
\begin{enumerate}
    \item $O$ commutes with every $g$ in $\{ g_i \}$, then $O$ is a stabilizer and no change is required;
    \item $O$ anticommutes with a single $g_i$, then there is a random measurement outcome $\sigma = \pm 1$ and $g_i$ is replaced with $\sigma O$;
    \item $O$ anticommutes with more than one $g_i$, where the set of anticommuting $g_i$s is denoted as $\{ g_1,...g_k\}$, then we perform a gauge transformation $g_j' = g_1 g_j$ for $2 \leq j \leq k$, and reduce to the previous case (as $O$ only anticommutes with $g_1$).
\end{enumerate}
The density matrix for a stabilizer state takes the form
\begin{align}
    \rho = \prod_{i = 1}^{N-S} \frac{\mathbb{I} + s_i}{2^N},
\end{align}
where $\{s_i\}$ is a list of stabilizer generators and $S$ is the entropy of the state.
A set of $N$ independent stabilizers fully specifies a pure state, giving $S = 0$. 
An incomplete list of generators, i.e. one with fewer than $N$ elements, defines a mixed state, with any ``missing'' generator contributing one bit of entropy. 
The completely mixed state is defined by an empty list ($S = N$), and its density matrix takes the form
\begin{align}
    \rho = \frac{\mathds{I}}{2^N}.
\end{align}
For time evolution of a mixed state, the update rules listed above must be modified in one case. 
Namely, if
\begin{enumerate}
    \item $O$ commutes with all the $\{g_i\}$ and
    \begin{enumerate}
        \item $O$ is a member of the stabilizer group, then $O$ is a stabilizer and no change is required;
        \item $O$ is \textit{not} a member of the stabilizer group, then $O$ is a logical operator. In this case, the measurement outcome is random ($\sigma = \pm 1$), a single bit of entropy is lost, and $\sigma O$ is added to the stabilizer group.
    \end{enumerate}
\end{enumerate}

\section{Review of Stabilizer and Subsystem Codes\label{app:subsystem_codes}}
\subsection{Stabilizer codes}
In this section, we provide further background on stabilizer codes. A \textit{stabilizer code} is one in which the states in code space are $+1$ eigenvalues of a set of stabilizer operators (see Appendix~\ref{app:stabilizer_codes} above). As an example, the three qubit repetition code has a two-dimensional code space spanned by the following logical states:
\begin{align}
    \ket{\bar{0}} &= \ket{000}, \\
    \ket{\bar{1}} &= \ket{111}.
\end{align}
The subspace can also be specified as the simultaneous $+1$ eigenspace of the following operators:
\begin{align}
    \mathcal{S} = \langle Z_1 Z_2, Z_2 Z_3 \rangle \;,
\end{align}
where $\langle \cdots \rangle$ denotes the group generated by those operators, which is Abelian. 
The code implicitly specifies {\it logical operators} $\bar{Z}$, $\bar{X}$ which act on the logical space as 
$\bar{Z}\ket{\bar{0}} = \ket{\bar{0}}$, 
$\bar{Z}\ket{\bar{1}} = - \ket{\bar{1}}$, 
$\bar{X}\ket{\bar{0}} = \ket{\bar{1}}$, 
$\bar{X}\ket{\bar{1}} = \ket{\bar{0}}$.
In this example, $\bar{Z} = Z_1$, $\bar{X} = X_1 X_2 X_3$ (up to multiplication by elements of $\mathcal{S})$.
Logical operators commute with $\mathcal{S}$ but are not part of it.

Errors are represented by arbitrary operators $E$ acting on the Hilbert space; however, by measuring the stabizers, this continuous set of possible errors effectively reduces to a discrete set (bit and phase flips on subsets of qubits), which either commute or anticommute with the stabilizers.
Measuring all stabilizers yields a set of outcomes, dubbed the {\it syndrome};
as long as an error $E$ anti-commutes with at least one element of the stabilizer group, it will give rise to a nontrivial syndrome and thus be detected.
The problem then becomes associating each possible syndrome to an error, in order to determine the best correction operation~\cite{shor_scheme_1995, gottesman_class_1996, bombin_topological_2010}.
Each stabilizer code has a {\it distance} $d$, defined as the minimum size (i.e., number of non-identity Pauli matrices) of a logical operator. A stabilizer code of distance $d = 2t+1$ can detect $2t$ errors and successfully correct $t$ errors. 
\subsection{Subsystem Codes}
Next, we move on to discussing subsystem codes~\cite{bacon_operator_2006, poulin_stabilizer_2005, aliferis_subsystem_2007}. A subsystem code is defined from a subgroup $\mathcal{G}$ of the $N$-qubit Pauli group $\mathcal{P}_N$, 
known as the \textit{gauge group}. In turn this is generated by a set of local operators $\mathcal{G} = \langle g_i \rangle$ known as ``checks'', the idea being that (like the stabilizers in a stabilizer code) these are operators we can measure in order to check whether errors have occurred. However, $\mathcal{G}$ is not an Abelian group (i.e., the checks are allowed to anticommute).
Given $\mathcal{G}$, we designate the stabilizer group $\mathcal{S}$ as the \textit{center} of $\mathcal{G}$, i.e. the subgroup of $\mathcal{G}$ containing elements that commute with all the members of $\mathcal{G}$. 
We may quotient the gauge group $\mathcal{G}$ by the stabilizer subgroup $\mathcal{S}$, to obtain a group $\mathcal{Q} \equiv \mathcal{G}/\mathcal{S}$ which is isomorphic to the Pauli group $\mathcal{P}_{n_G}$ on a number $n_G$ of qubits; this defines a set of ``gauge qubits''.
Finally, logical operators $\mathcal{L}$ are Pauli strings that commute with $\mathcal{G}$, defined up to stabilizers (formally, we have $\mathcal{L} = \mathcal{N}(\mathcal{G}) / \mathcal{S}$, where $\mathcal{N}$ denotes the {\it normalizer} of the gauge group $\mathcal{G}$ in $\mathcal{P}_N$ and we are taking a quotient by the normal subgroup $\mathcal{S}$~\cite{bombin_topological_2010}). 

One helpful way to think of this construction is as a conventional stabilizer code, specified by $\mathcal{S}$, with a code subspace that has a product structure $A \otimes B$ (hence ``subsystem''); the two factor Hilbert spaces describe logical qubits ($A$) and gauge qubits ($B$).
However, this code space is redundant, as we only aim to correct errors on the logical qubits, i.e. on the subsystem $A$. Thus we are allowed to dump errors on the gauge qubits during the correction process: such errors affect only the subsystem $B$, and can be regarded as gauge transformations. 
This may prove advantageous by allowing the measurement of lower-weight operators.
While the checks do not commute, they can be measured sequentially in order to effectively measure generators of the stabilizer group $\mathcal{S}$.
\subsection{Counting of Logical Qubits}
Any set of Pauli strings $\{g_i \}$ may be used to define a subsystem code, by taking $\mathcal{G} = \langle\{g_i\} \cup \{i\mathbb{I}\} \rangle$, identifying the stabilizer group $\mathcal{S}$ as its center and the gauge qubits from the quotient $\mathcal{G}/\mathcal{S}$, and finally the logical group $\mathcal{L}$ as the group of Pauli strings that commute with $\mathcal{G}$, modulo stabilizers. 
Letting $n_S = \text{dim}(\mathcal{S})$ be the number of independent stabilizer generators, 
$n_G = \frac{1}{2} \text{dim}(\mathcal{G})$ be the number of gauge qubits (note each gauge qubit contributes a pair of independent anticommuting operators playing the role of $X$ and $Z$), and $n_L = \frac{1}{2} \text{dim}(\mathcal{L})$ be the number of logical qubits, by fixing the Hilbert space dimension (i.e. counting the total number of qubits) we must have $N = n_S + n_G + n_L$. 
From the set of checks $\{g_i\}$ we can determine $n_S$ and $n_G$ (noting also that $\text{dim}(\mathcal{G}) = n_S + 2n_G$), and thus the number of logical qubits $n_L = N - n_S - n_G$.

The honeycomb model studied in this paper, when treated as a subsystem code with checks given by the Kitaev Hamiltonian terms $\mathcal{C}$, stores {\it no} logical qubits: $n_L = 0$. 
For an $L\times L$ honeycomb lattice, there are $N = 2L^2$ physical qubits. Every physical qubit belongs to 3 links, giving $\frac{3}{2}N$ links, where the factor of $\frac{1}{2}$ comes from double counting. Every link is a check, and since the product of all checks is the identity (and this is the only such algebraic constraint), we have that $\text{dim}(\mathcal{G}) = \frac{3}{2}N - 1$. 
Products of checks on cycles form the stabilizers. These cycles can be either topologically trivial or nontrivial. 
The simplest topologically trivial cycles are the 6-site plaquette operators, which can be used to generate any contractible cycle. For the $L\times L$ lattice, there are $L^2 = N/2$ plaquettes, however the product of all plaquettes is the identity. In addition, there are two topologically nontrivial cycles (that wrap around the torus in the two directions). We then have the size of the stabilizer group: $n_S = N/2+1$. 
The number of gauge qubits obeys $n_S + 2n_G = \text{dim}(\mathcal{G})$, and thus $n_G = \frac{1}{2} (N-2)= N/2 -1$. 
We see therefore that the $N$ physical qubits split into $N/2+1$ stabilizers and $N/2-1$ gauge qubits, leaving no logical qubits: $n_L = N - n_S -n_G = 0$.

In the Floquet code implementation~\cite{hastings_dynamically_2021}, the two topologically-nontrivial stabilizers are {\it never} generated, even though they are nominally part of $\mathcal{S}$; thus one has $n_S^{\rm eff} = n_S - 2 = N/2 - 1$, and $n_L^{\rm eff} = 2$. In other words, the topological stabilizers play the role of logical operators, namely the two commuting logical $\bar{Z}$'s (dubbed ``inner'' logicals). Their conjugate $\bar{X}$'s (``outer'' logicals) are not in $\mathcal{G}$.

\section{Critical contribution to the tripartite mutual information \label{app:I3}}

Here we explain the origin of the correction to the tripartite mutual information $\mathcal{I}_3$ in addition to the topological term observed in Fig.~\ref{fig:I3}, and argue that it is a signature of a critical phase.

Let us consider the geometry displayed in the inset to Fig.~\ref{fig:I3}(a)---three contiguous regions $A$, $B$ and $C$ such that $ABC$ forms a doughnut-shaped region. Further, let $D$ denote the complement of $ABC$, and let $D = D_1\cup D_2$ with $D_1$ the interior region (surrounded by $ABC$) and $D_2$ the exterior region.
We assume that the instantaneous stabilizer group for the state admits a set of generators that are products of checks along paths on the lattice (this will almost always be true at sufficiently late times, regardless of the initial state). 
These comprise (i) plaquette stabilizers, (ii) topological stabilizers (products of checks along the two noncontractible loops of the torus), and (iii) open strings, with endpoints at two sites $i$ and $j$ on the lattice. Note that the latter are specified only from their endpoints $i$ and $j$---the path connecting the endpoints may be deformed at will by multiplying with plaquette stabilizers.

We now consider the tripartite mutual information $\mathcal{I}_3(A:B:C)$ of a state in the form described above, in the thermodynamic limit $L \to \infty$. Note from Fig. ~\ref{fig:I3}(a) that $L$ is the system size and the region containing $A,B$ and $C$ is a proportionate fraction of $L$. From the definition $\mathcal{I}_3(A:B:C) = S_A + S_B + S_C - S_{AB} - S_{BC} - S_{CA} + S_{ABC}$, we know that area-law terms cancel out (as each boundary appears in the sum twice with a positive sign and twice with a negative sign).
Similarly, we know that the topological terms add up to $-2$~\cite{kitaev_topological_2006, yao_entanglement_2010}.
Thus it remains to consider the effect of open strings.

Recall that the entanglement entropy of some region $R$ is given by the difference between its volume (number of qubits) and the number of independent stabilizers contained entirely within $R$. Thus each open string with an endpoint inside $R$ and the other outside of it contributes $+1$ bit to the entropy $S_R$. Further, if $R$ is not contiguous, open strings with endpoints in distinct connected components of $R$ may also contribute $+1$ bit (as any string operator connecting the endpoints must necessarily exit $R$).
Letting $n_{RR'}$ be the number of open strings with an endpoint in region $R$ and the other in region $R'$, we have:
\begin{itemize}
    \item $S_A = n_{AB} + n_{AC} + n_{AD} + \dots $, where $\dots$ stands for short-range and topological contributions. $S_B$ and $S_C$ have analogous expressions (obtained by permuting $A$, $B$ and $C$).
    \item $S_{AB} = n_{AC} + n_{BC} + n_{AD} + n_{BD} + \dots$, and analogous expressions for $S_{BC}$ and $S_{CA}$.
    \item $S_{ABC} = S_D = n_{AD} + n_{BD} + n_{CD} + \Theta(n_{D_1 D_2}) + \dots$, where $\Theta(n)=1$ if $n>0$, $\Theta(n)=0$ otherwise. 
\end{itemize}
Note that, as $D$ is non-contiguous, open strings with endpoints in the two distinct connected components $D_1$ and $D_2$ are {\it not} entirely contained in $D$.
However, any two such strings can be smoothly deformed (by multiplication with plaquette stabilizers) in such a way that they perfectly overlap everywhere in $ABC$; thus any number $n_{D_1 D_2}>0$ of such strings only contributes one bit of entropy~\cite{lavasani_monitored_2022}. This accounts for the contribution $\Theta(n_{D_1 D_2})$ in the last item.

Adding all contributions yields
\begin{equation}
    \mathcal{I}_3(A:B:C) = 2S_\text{topo} + \Theta( n_{D_1 D_2}(L) )\;,
    \label{eq:I3topo}
\end{equation}
and thus, upon disorder averaging,
\begin{equation}
    \delta \mathcal{I}_3 = \textsf{Prob}(n_{D_1D_2}(L) \geq 1)
    \label{eq:app_deltaI3}
\end{equation}
($\delta \mathcal{I}_3 \equiv \mathcal{I}_3 - 2S_{\rm topo}$ is the critical contribution to the mutual information).
In the $L\to \infty$ limit, while $A,B,C$ are still held as finite fractions of $L$, the second term in ~\ref{eq:I3topo} depends on the distribution of lengths for open-string stabilizers, $P(\ell)$. The counting of $n_{D_1D_2}$ is given by
\begin{equation}
    n_{D_1 D_2}(L) = \int_{D_1} dr_1 \int_{D_2} dr_2 \frac{1}{|\vec{r}_1-r_2|} \cdot P(|r_1-r_2|).
    \label{eq:nd1d2}
\end{equation}
Here we are integrating over all possible endpoints, weighted by the probability of a stabilizer connecting them, to count the number of stabilizers that begin in $D_1$ and end in $D_2$. 
If $P(\ell)$ is short-ranged, with a finite characteristic length $\xi$, as is expected in the area-law phase, then Eq.~\eqref{eq:nd1d2} yields $n_{D_1D_2} \sim {\rm poly}(L) e^{-L/\xi}$, which vanishes in the $L \to \infty$ limit, and we recover the expected topological term.

In the critical phase we expect a long-ranged distribution $P(\ell) \sim \ell^{-2}$, based on the multiplicative violation of the area-law~\cite{nahum_entanglement_2020}.
However this scaling holds for $\ell \ll L$, while $n_{D_1 D_2}$ depends on the probability of strings of length $\approx L/2$. It is thus not clear {\it a priori} what to expect.

\begin{figure}
\centering
\includegraphics[width=\columnwidth]{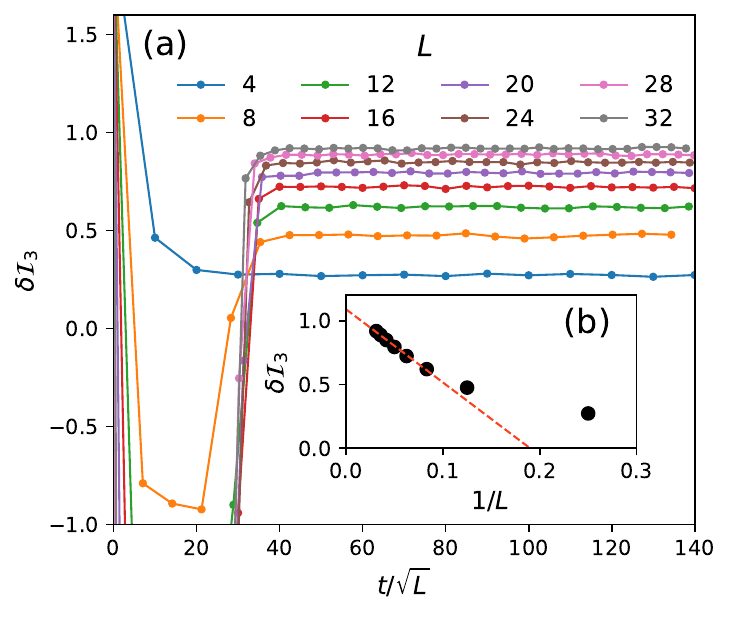}
\caption{Scaling of the tripartite mutual information in the critical phase, $p_x = p_y = p_z = 1/3$, for the topologically nontrivial geometry of subsystems used in Fig.~\ref{fig:I3}. 
(a) Non-topological term $\delta\mathcal{I}_3 \equiv \mathcal{I}_3 +2$ as a function of rescaled time $t/\sqrt{L}$. A pleateau is achieved at $t\approx 30\sqrt{L}$; the height of the plateau weakly increases with system size. Data averaged over between $4\times 10^3$ and $10^4$ realizations depending on size.
(b) Plateau value of $\delta \mathcal{I}_3$ (average of datapoints with $t/\sqrt{L}\geq 50$) vs $1/L$. Extrapolation to the $L\to\infty$ limit (dashed line) indicates a finite value close to $+1$.
\label{fig:deltaI3_center}}
\end{figure}

Numerics on the central point of the critical phase, $p_x = p_y = p_z = 1/3$, Fig.~\ref{fig:deltaI3_center}(a), shows an asymptotic value of $\delta\mathcal{I}_3$ that grows weakly with system size and appears to saturate to a constant, $\delta\mathcal{I}_3 \sim \text{const.} + o(1)$. 
In Fig.~\ref{fig:deltaI3_center}(b) we see that, by choosing the $o(1)$ finite-size correction to be $\propto 1/L$, the extrapolated $L\to\infty$ limit is close to 1.
A value of $+1$ would indicate that $n_{D_1 D_2}(L) \geq 1$ with unit probability, see Eq.~\eqref{eq:app_deltaI3}; the fact that the extrapolated value slightly exceeds $+1$ could be due to the choice of extrapolation or slow temporal drift of the $\delta{I}_3(t)$ curves beyond the observed time scales.
Furthermore, we note that the plateau value of $\delta\mathcal{I}_3$ is achieved at time $t \propto \sqrt{L}$, which is another manifestation of the $z = 1/2$ dynamical behavior identified in Sec.~\ref{sec:crit_purif}.

For the geometry considered in Fig.~\ref{fig:I3_topo}, the region $D_1$ vanishes, and thus the critical contribution $\Theta(n_{D_1 D_2})$ also vanishes. Further, the topological entanglement entropy is counted on net only once, giving $\mathcal{I}_3 \to S_\text{topo} = -1$ in both phases, as is seen in Fig.~\ref{fig:I3_topo}.

\section{Purification in the Critical Phase \label{app:purification_times}}

\begin{figure}
    \centering
    \includegraphics[width = \columnwidth]{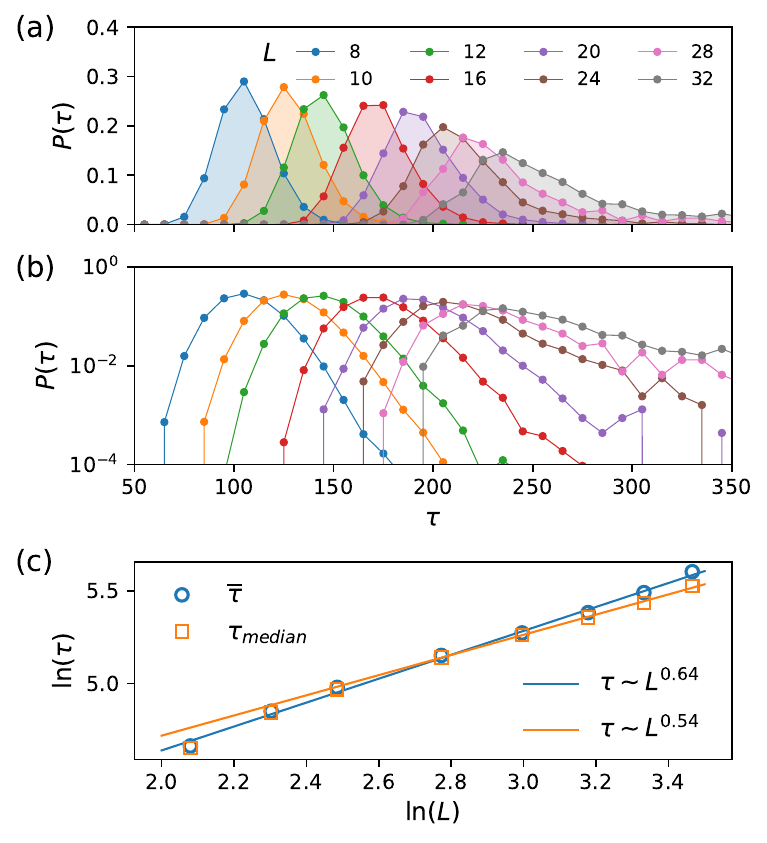}
    \caption{Distribution of purification times in the critical phase, $p_x = p_y = p_z = 1/3$. 
    (a) Distribution $P(\tau)$ of purification times $\tau$ over realizations ($10^3$ to $10^5$ total realizations depending on size) for different system sizes.
    (b) Same data in semilogarithmic scale; broadening towards large $\tau$ is visible at the larger system sizes.
    (c) Average and median purification time vs system size $L$. A discrepancy between the two is visible at the larger sizes.
    Fitting a power law to the $L>12$ datapoints yields dynamical exponent $z$ consistent with $1/2$ for the median time, and slightly larger for the average time. 
    }
    \label{fig:dist_timings}
\end{figure}

In this section, we discuss dynamical purification in the critical phase.
The data in Fig.~\ref{fig:purification_dynamics} in the main text shows that the entropy of an initially-mixed state decays as $S(t) \sim e^{-t/L^z}$ at late times, with $z = 1/2$. 
Here present additional data on this scaling by focusing on the distribution of purification times over realizations of the dynamics. 

Because the state of the system is a stabilizer state at all times, the entropy takes on only integer values, and the state becomes exactly pure at a specific time in each run.
Thus we can sharply define a purification time as $\tau \equiv \min\{\tau :\ S(\tau )=0\}$. 
In Fig.~\ref{fig:dist_timings}(a), we show the distribution of times $\{\tau \}$ over random realizations of the measurement-only dynamics at the center of the phase diagram (and of the critical phase), $p_x = p_y = p_z = 1/3$. We note that the distribution broadens with increasing system size.
The {\it median} purification time $\tau_\text{med} = \text{median}\{\tau\}$, shown as a function of system size in Fig.~\ref{fig:dist_timings}(b), scales as $\tau_\text{med} \propto L^z$ with $z = 0.52 \simeq 1/2$. 

However, the {\it mean} purification time $\overline{\tau}$ deviates significantly from this trend: the best fit at the available sizes yields $z = 0.60$, but the exponent is clearly drifting towards a larger value.
The mean is thus dominated by the tails of the distribution, and the relative weight of the tails increases with system size.

The physical origin of these tails (representing abnormally long-lived qubits at late time in the dynamical purification) is the $z = 2$ critical dynamics of the fermionic sector once the flux sector is completely purified~\cite{lavasani_monitored_2022}. In other words, in some realizations (which are rare at these sizes, but become more common with increasing $L$), the plaquette fluxes completely purify while certain open strings (fermions) remain mixed; the subsequent dynamics of the fermions on a completely-purified flux background is well understood analytically~\cite{lavasani_monitored_2022} and is much slower ($z=2$) than the previous dynamics, causing tails in the distribution of $\{\tau\}$.
To understand whether these tails become ultimately dominant in the $L\to\infty$ limit, it would be necessary to study the dynamics of the fermions in an incompletely-purified flux background (i.e., one where not all plaquette fluxes are in the stabilizer group, and thus fermionic strings cannot be deformed arbitrarily). This is an interesting problem for future work.

\section{1D-2D Critical Crossover}
\begin{figure}

    \centering
    \includegraphics[width = \columnwidth]{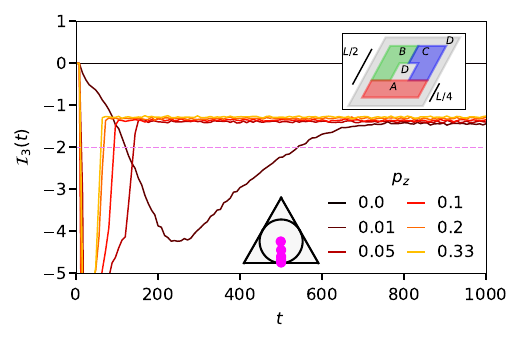}
    \caption{Tripartite mutual information shown for the 1D critical point and the 2D critical phase. We used a system size of $L=16$ here. Insets show the points at which the parameters were sampled as well as the geometry used for the $\mathcal{I}_3$ calculation.}
    \label{fig:1D2DI3}
\end{figure}

\begin{figure}
    \centering
    \includegraphics[width = \columnwidth]{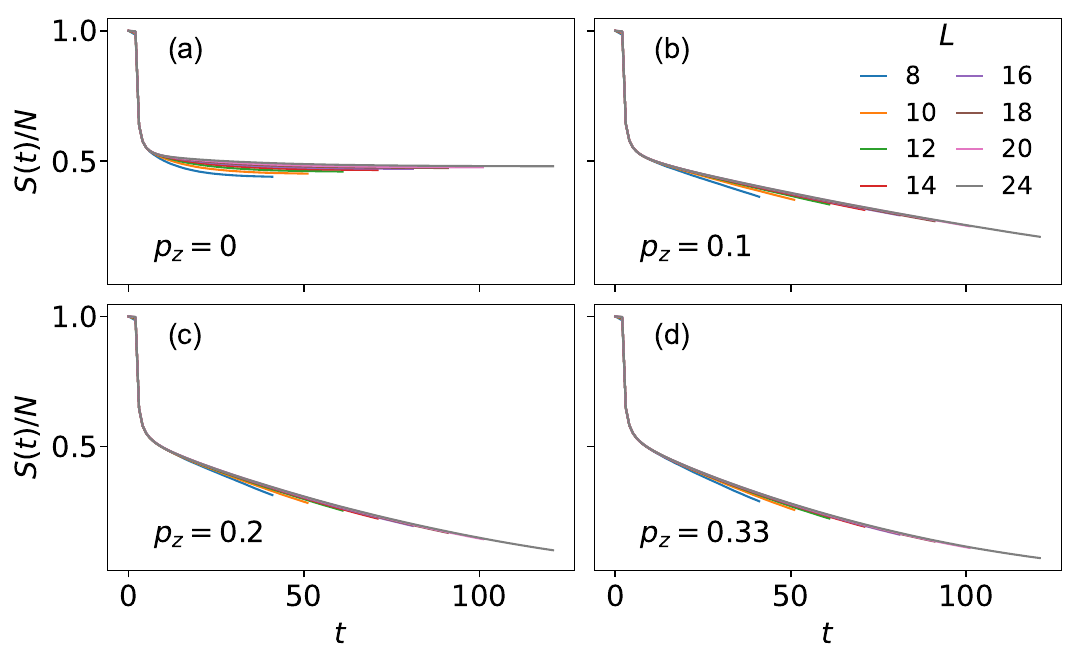}
    \caption{Mixed state evolution near the 1D-2D critical transition, where (a),(b),(c),(d) are $p_z=0,0.1,0.2,0.33$ respectively. We see that at $p_z = 0$, the system plateaus to a constant purity, and for any nonzero $p_z$, the purity decreases past this plateau.}
    \label{fig:1D2DI3}
\end{figure}

In the main text, we primarily focused on the transition between the area-law phase and the critical phase that occurs as $p_z$ is decreased from $1 \to \frac{1}{3}$. Here we study the dynamics of the model as $p_z$ is increased from $0 \to \frac{1}{3}$. As stated earlier, along the boundaries of the phase diagram the system decouples into wires (at the vertices of the phase diagram, it decouples into dimers). In particular, the 1D wires evolve under two types of non-commuting measurements that define a bipartite ensemble. The properties of this dynamics are studied e.g. in Refs.~\cite{nahum_entanglement_2020, ippoliti_entanglement_2021}. For $p_z=0$ each 1D wire is at a critical point of this dynamics.
Any nonzero $p_z$ couples the wires and makes the dynamics genuinely two-dimensional, leading to a dramatic difference. One can observe this in the tripartite mutual information, calculated for the geometry in Fig \ref{fig:I3}(a). As shown in the main text, both the area-law and critical phases are topological, giving a negative value of $\mathcal{I}_3$. However, when the system decouples into 1D wires, it becomes non-topological, and as expected we numerically observe a value of $\mathcal{I}_3 = 0$. 
As a practical matter, note that in the computation of $\mathcal{I}_3$ we take $L$ to be a multiple of 8. Otherwise, for this choice of $A,B,C$ regions, a 1D wire overlaps all three regions and may give rise to spurious nonzero contributions to $\mathcal{I}_3$.

A further difference between the 1D and 2D criticality is that, while the 1D critical point is known to admit a CFT description in terms of loop percolation~\cite{nahum_entanglement_2020}, we showed in the main text that this is not the case for the 2D critical phase in this model.

In addition, we study dynamical purification of an initially-mixed state in the model. At the boundary of the phase diagram ($p_z = 0$), the entropy density drops rapidly but then plateaus at a finite value. This is due to the fact that, in the absence of $ZZ$ measurements, certain stabilizers cannot be generated; namely, $YY$ on an $X$-type bond and $XX$ on a $Y$-type bond act as ``logical'' operators, that commute with all measurements but remain un-measured throughout the dynamics. 
This sets a finite floor to the system's entropy density. For $p_z > 0$, we instead see that the entropy density continues to decrease past the initial drop. As one moves closer to the center of phase diagram, this decrease in entropy density becomes more rapid.

When $p_z = 0$, we find that this plateau occurs slightly below an entropy density $S/N = 1/2$, and as $N\to \infty$, the plateau approaches $1/2$. To understand this, consider an example 1D system of four qubits, with periodic boundary conditions.
\begin{align}
    \begin{tikzpicture}
    \filldraw (0,0) circle (3pt);
    \filldraw (1,0) circle (3pt);
    \filldraw (1,-1) circle (3pt);
    \filldraw (0,-1) circle (3pt);
    \draw [color=red] (0.2,0)--(.8,0);
    \draw [color=blue](1,-0.2)--(1,-.8);
    \draw [color=blue](0,-0.2)--(0,-.8);
    \draw [color=red](0.2,-1)--(.8,-1);
    \node at (.5,.5) {$YY$};
    \node at (1.5,-.5) {$XX$};
    \end{tikzpicture}
\end{align}

If one first measures the $YY$ bonds starting from a completely mixed state, the two $YY$ bonds will generate the stabilizer group. 
However, if one then measures $XX$ bonds, a set of generators for the stabilizer group will now consist of the $XX$ bonds and the product of the two previously-measured $YY$ bonds, that is, $YYYY$. It is straightforward to see that a fourth independent generator cannot be added by only measuring these checks, thus the final value of entropy is 1 bit. 
In this instance we thus have $3$ stabilizers. By extension, for a chain of length $L$ with alternating anti-commuting bonds (e.g. $XX$ and $YY$), the generators of the stabilizer group can at most consist of all the bonds of one type (e.g. $X_{2i} X_{2i+1}$), and the product of all the bonds of the other type (e.g. $Y_1Y_2\cdots Y_L$). This yields a minimum value of entropy $S \geq \frac{L}{2} - 1$. In the honeycomb, there are $L$ such chains and so we have that
\begin{align}
    S \geq \frac{N}{2} - L.
\end{align}

\bibliography{main}

\end{document}